\documentclass[a4paper,11pt]{article}
\usepackage{pos}
\usepackage{lineno}
\title{Muon number rescaling in simulations of air showers}
 \ShortTitle{Muon number rescaling...}

\author*[a]{Dariusz Góra}
\author[a]{Nataliia Borodai}
\author[b]{Ralph Engel}
\author[b]{Tanguy Pierog}
\author[a]{Jan Pękala}
\author[b]{Markus Roth}
\author[a]{Jarosław  Stasielak}
\author[b]{Michael Unger}
\author[b]{Darko Veberic}
\author[b]{Henryk Wilczy\'nski}

\affiliation[a]{Institute of Nuclear Physics PAN,\\
  Radzikowskiego 152, Cracow, Poland}

\affiliation[b]{Karlsruhe Institute of Technology (KIT),\\
Institute for Astroparticle Physics, Karlsruhe, Germany}

\emailAdd{Dariusz.Gora@ifj.edu.pl}
\abstract{The number of muons in extensive air showers predicted using LHC-tuned hadronic interaction models, such as EPOS-LHC and QGSJetII-04, is smaller than observed in showers recorded by leading cosmic ray experiments. In this paper, we present a new method to derive muon rescaling factors by analyzing reconstructions of simulated showers. The z-variable used (difference of initially simulated and reconstructed total signal in detectors) is connected to the muon signal and is roughly independent of the zenith angle but depends on the mass of primary cosmic ray. The performance of the method is tested using Monte Carlo shower simulations for the hybrid detector of the Pierre Auger Observatory. Having an individual z-value from each simulated hybrid event, the corresponding signal at 1000 m from the shower axis, and using a parametrization of the muon fraction in simulated showers, we can calculate the multiplicative rescaling parameters of the muon signals in the ground detector even for an individual event. We can also  study its dependence as a function of zenith angle and the mass of primary cosmic ray. This gives a possibility not only to test/calibrate the hadronic interaction models, but also to derive the $\beta$-exponent, describing an increase of the number of muons as a function of primary energy and mass of the cosmic ray. Detailed simulations show dependence of the $\beta$-exponent on hadronic interaction properties, thus the determination of this parameter is important for understanding the muon deficit problem.}

\FullConference{37$^{\rm{th}}$ International Cosmic Ray Conference (ICRC 2021)\\
		July 12th -- 23rd, 2021\\
		Online -- Berlin, Germany}

\begin{document}
\maketitle

\vspace{-0.5cm} 
\section{Introduction}
\vspace{-0.3cm}
Simulations of extensive air showers using current hadronic interaction models predict too small number of muons, which is known as the muon deficit problem. The muon number predicted by the LHC-tuned  models, such as EPOS-LHC~\cite{epos} and QGSJetII-04~\cite{qgsjet}, is 30\% to 60\% lower than what is observed at the shower energy of $10^{19}$ eV~\cite{allen}. Since data interpretation relies on simulations, the problem with muons has deep implications: the data suggest a much heavier composition of cosmic rays based on muons only than the composition derived from $X_{\mathrm{max}}$~\footnote{The $X_{\mathrm{max}}$ is the atmospheric depth at which the longitudinal development of an air shower reaches the maximum number of particles.} measurements~\cite{xmax}. 
%\begin{figure}
%\centering
%\includegraphics[width=0.95\textwidth,height=0.45\textwidth]{TD-sim.png}
%\caption{Simulated hybrid event with energy $10^{19}$ eV seen by the fluorescence and surface detector of the %Pierre Auger Observatory. Left: The reconstructed longitudinal profile (blue) of an illustrative air shower with %its matching simulated shower (red solid), using QGSJet-II-04 for proton. Right: The simulated ground signals %for the same event (p: blue full circles) in units of vertical equivalent muons (VEM)~\cite{vem}; curve is the %lateral distribution function (LDF) fit to the signal. Dashed red lines indicates an energy estimator at 1000 %meter from the shower core.}
%\label{fig1}
%\end{figure}
To study the muon number problem, a top-down (TD) reconstruction method was proposed~\cite{allen}. The main aim of the TD reconstruction is to predict signals in the fluorescence (FD) and surface detectors (SD) of the Pierre Auger Observatory~\cite{auger} (Auger) on a simulation basis. Therefore, the TD-method finds a simulated shower, which has a particle distribution of electromagnetic component along the shower axis (longitudinal profile) similar to the longitudinal profile of the data shower (a reference profile). The reference longitudinal profile is linked with the electromagnetic component of the shower, so the method relies on the fact that this part is accurately simulated. As an output, the TD-method provides a reconstructed event, in which the signals in detectors are determined using Monte Carlo (MC) simulations. The simulated SD signals in the output shower, which depend on the interaction models, may be then compared with the data. The SD signals also include  the contribution of muons, which are tracers of properties of the hadronic interactions. Comparison of simulated SD signals with the corresponding signals in the data shower provides an opportunity to check the correctness of lateral distributions of the simulated showers. Since the lateral distributions are sensitive to the hadronic interaction models, an analysis of these distributions provides an opportunity to investigate indirectly the interaction models at energies above  energies, at which the accelerator data are available. So it is expected that the TD-method will allow us to calibrate the interaction models, and to reduce discrepancy between the data and simulations.

In this paper, we present a new method for determining muon scaling factors by analyzing reconstructions of simulated showers, i.e. instead of a real shower, a simulated one is used as the reference event (MOCK-DATA). The $z$-variable used (the difference between the initially simulated and reconstructed total signal at the detectors) is related to the muon signal and is approximately independent of the zenith angle, but depends on the mass of the primary cosmic ray. The performance of the method is tested using MC simulations for the hybrid detector of the Pierre Auger Observatory.
\begin{figure}
\vspace{-0.7cm}
\centering
\includegraphics[width=0.52\textwidth,height=0.30\textwidth]{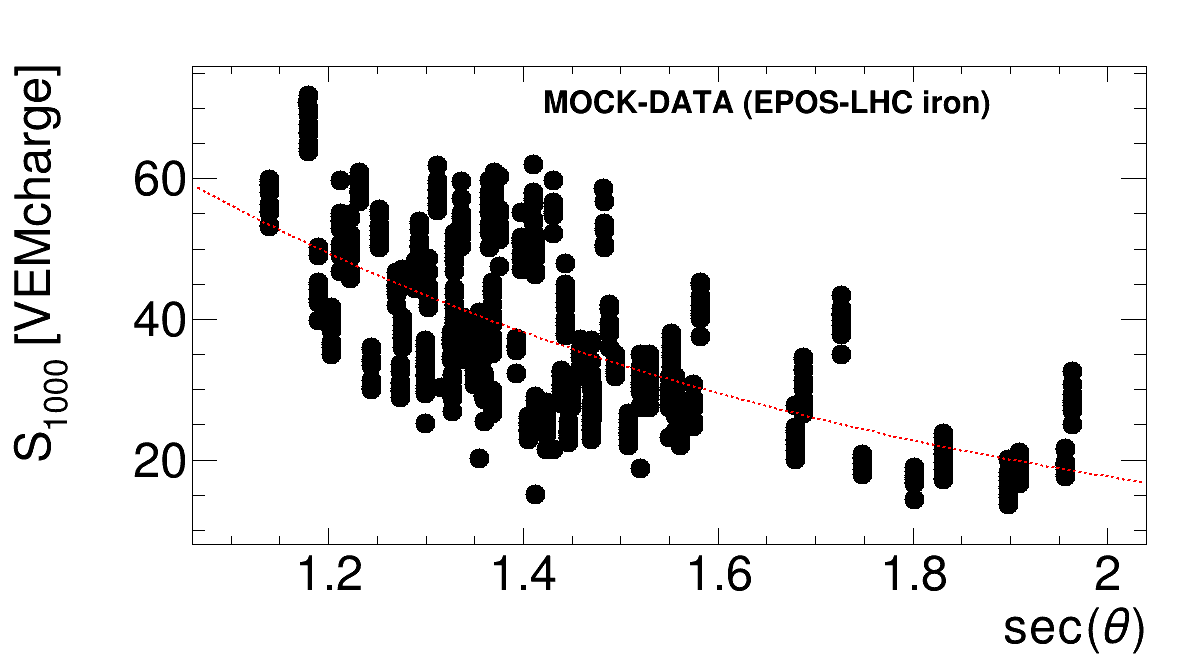}
\vspace{-0.2cm}
\caption{The total MC signal, $S_{1000}$  as a function of zenith angle $\theta$, for the SD station at a distance 1 km from the shower core, $10^{19}$ eV iron induced air showers and EPOS-LHC model. The total signal in units of vertical equivalent muons (VEM~\cite{auger}) varies between $~\sim 20$ VEM and $\sim 60$ VEM.}
\label{fig2}
\vspace{-0.3cm}
\end{figure}
\vspace{-0.3cm} 
\section{Description of the Top-Down scheme and z-method}
\vspace{-0.3cm}
The top-down reconstruction chain used in this work is an improved version of the
reconstruction chain described in~\cite{allen}. The main improvement is the use of CORSIKA~\cite{corsica} as the main simulation tool, instead of SENECA software~\cite{seneca}. CORSIKA was developed intensively during recent years, so when compared with  SENECA,
it currently provides more reliable and accurate simulations. All of the improvements of the method are described in detail in~\cite{tiff}. In this work, the TD-chain includes a simulation of the surface detector response for the CORSIKA simulated event (reference shower). The Auger observatory response for the reference  shower is  simulated in the hybrid mode (the event is seen by SD and FD), using the $\overline{\mathrm{Off}}$\underline{line} software~\cite{offline}, which provides  20 reconstructions of the event. From them 10 reconstructions are selected for comparison  of the station signals with the reference MC event. The selection criterion  is the  value of $\chi^2$ calculated between the reference MC  and reconstructed longitudinal profiles,  for more details see~\cite{tiff}.  So 10 final reconstructions of the event, with the lowest values of the $\chi^2$, are used in subsequent SD signal analysis~\footnote{We use 78 reference events (i.e., 780 $\overline{\mathrm{Off}}$\underline{line} reconstructions) for EPOS-LHC and 89 (i.e., 890) for QGSJetII-04.}. The TD-chain  was performed independently for two interaction models (EPOS-LHC and QGSJetII-04), and for 4 types of primary particle: proton, helium, nitrogen and iron.

To take into account the muon excess seen by  Auger and for illustration of the method, the iron EPOS-LHC  simulations are used as the MOCK-DATA set, see Figure~\ref{fig2}. Note that the total SD signal for EPOS-LHC in the case of iron simulations is quite similar to the SD signal measured by  Auger at energy $10^{19}$ eV~\cite{balaz}. On the other hand, to reconstruct the total muon signal in the MOCK-DATA set we  use MC simulations from QGSJetII-04.

Shower particles are usually classified into four components: the muonic component, 
the electromagnetic component  stemming from muon interactions and muon decays, the purely electromagnetic component, and the  electromagnetic component from low-energy hadrons~\cite{signal}. In a first approach, the contribution  from the electromagnetic muon halo  and from the hadron jet component  can be neglected ~\cite{signal}, so that the main contribution to the total  ground signal at 1000 m ($S_{1000}$) comes from the purely electromagnetic and muonic components~\cite{allen}.
\begin{figure}
\vspace{-0.1cm}
\centering
\includegraphics[width=0.48\textwidth,height=0.30\textwidth]{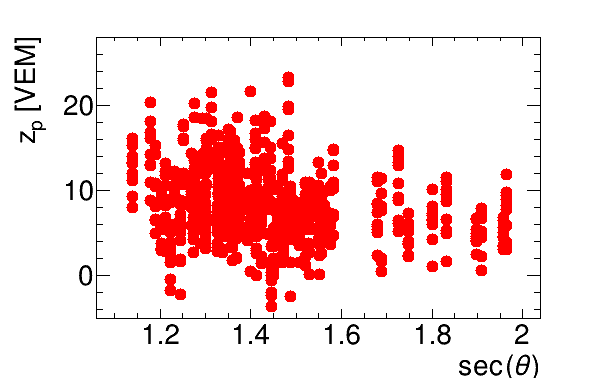}
\includegraphics[width=0.48\textwidth,height=0.30\textwidth]{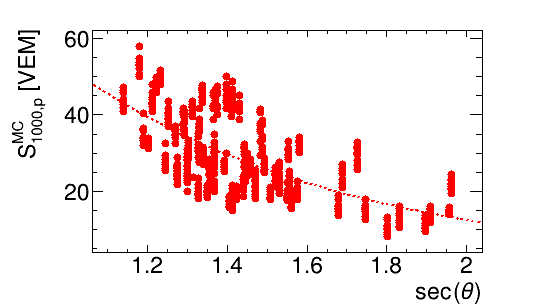}
\caption{Left: The $z_{p}$-variable (left) and the total MC signal at 1000 m (right) as a function of zenith angle (for proton air showers with energy $10^{19}$ eV obtained with QGSJetII-04).}
\label{fig3}
\vspace{-0.3cm}
\end{figure}
Thus, to a first approximation, these  contributions to the total signal at 1000 m can be described by two parts:  $S_{\mathrm{EM}}$ and $S_{\mu}$. 

As we noted above, the observed   SD signal  of ultra high energy  air showers is significantly larger than predicted by hadronic models tuned to fit the accelerator data~\cite{allen}.
Such disagreement can be  described  by introducing linear scaling factors: for the electromagnetic part  $R_{E}$ and the hadronic/muonic part: $R_{\mu}$. 
Following this approach for a single shower $j$, the  simulated  ground signal at 1000 m  from QGSjetII-04  MC and MOCK-DATA can  be written as:
\begin{eqnarray}
S_{1000,j}^{\mathrm{MC}} \equiv S_{\mathrm{EM,j}}^{\mathrm{MC}}+S_{\mu,j}^{\mathrm{MC}}, \hspace{9.1cm}\\
S^{\mathrm{MOCK-DATA}}_{1000}(R_E,R_{\mu})_{j} \equiv S_{\mathrm{EM,j}}^{\mathrm{MOCK-DATA}}+S_{\mu,j}^{\mathrm{MOCK-DATA}}= R_{E}S_{\mathrm{EM},j}^{\mathrm{MC}}+R_{\mu,j}R_{E}^{\alpha}S_{\mu,j}^{\mathrm{MC}}.
\end{eqnarray} 
In Eq.~(2) we have used the fact that some of the electromagnetic particles produced by muons in decay or radiation processes, as well as by low-energy $\pi^{0}$s, can be attributed to the electromagnetic signal by introducing an additional factor $R_{E}^{\alpha}$; but the muons that result from photoproduction are assigned to the electromagnetic signal, $S_{\mathrm{EM}}$~\footnote{The value of $\alpha$ is a prediction of hadron event generators and also depends on mass; in practice, both EPOS and QGSJetII-04 simulations find $\alpha \simeq 0.9$, relatively independent of the composition~\cite{alpha}.}. As shown in~\cite{allen},  no rescaling is needed for the electromagnetic part, where the most likely solution is $R_{E}=1$. Furthermore, in the TD-method the reference longitudinal profile is related to the electromagnetic component of the shower, so the method relies on the fact that this part is accurately simulated. Hence the assumption $R_{E}=1$ used in our analysis.

In this work, we used the difference between MOCK-DATA and the MC ground signal as the main observable, e.g. $z_j\equiv S_{1000,j}^{\mathrm{MOCK-DATA}}-S_{1000,j}^{\mathrm{MC}}$. This is because this variable is a natural indicator of the discrepancy between data and MC, and ideally the discrepancy should be zero. Other interesting feature arise from Eqs.~(1) and (2). For $R_E=1$  by simple subtraction we obtain  
\begin{equation}
S_{\mu,i,j}^{\mathrm{MC}}=\frac{z_{i,j}}{R_{\mu,i,j}-1}.
\end{equation}
  This is the key equation for the method presented in this paper. The formula  shows that the muon  MC  signal is proportional to the  difference  between data and MC signal, i.e. variable $z_{i,j}$, where 
 a proportionality coefficient  depends on  the muon scaling factor $R_{\mu,i,j}$.   Since  an average difference  $\langle z_{i} \rangle$  depend on the type of primary, we introduced scaling factors which  depend on mass  (additional  index $i  \in \{\mathrm{p,He, N, Fe}\}$ for $R_{\mu}$). Another argument for introducing different scaling factors for different primaries comes from the fact  that  the  average muon number is different for different primaries, i.e.  it increases with the atomic mass of primary particle. Therefore, 
  showers initiated by heavier primary with the same energy will  contain larger number of muons, and thus larger  ground muon signal is expected,  see for example~\cite{heitler}. 
  Finally, because $z_{i,j}$ is connected to the muon ground signal, this variable is roughly independent of the zenith angle. Indeed, as is shown for example in Figure~\ref{fig3} (left), the $z_p$-distribution only slightly depends on the zenith angle, in contrast to the total ground signal at distance 1000 m shown in Figure~\ref{fig3} (right). 
  
It is worth mentioning that having an individual value of $z_{i,j}$ with help of  MC simulations, the corresponding SD signal at 1000 m, and using the parametrization of the muon fraction, we can get the muon scaling factor even for an individual hybrid event $j$: 
\begin{equation}
R_{\mu,i,j}(\sec(\theta))=1+\frac{z_{i,j}(\sec(\theta))}{g_{\mu,i}(\theta) \times S^{\mathrm{MC}}_{1000,i,j}(\sec(\theta))},
\end{equation} 
where $S^{\mathrm{MC}}_{\mu,i,j}\equiv g_{\mu,i}(\theta) \times S^{MC}_{1000,i,j}$.
This approach provides an opportunity to study the dependence of the muon correction factor 
as a function of zenith angle, which may be useful for calibrating hadron interaction models.
%\begin{figure}
%\vspace{-0.5cm}
%\centering
%\includegraphics[width=0.49\textwidth,height=0.340\textwidth]{s_iron-epos_vs_cos.png}
%\includegraphics[width=0.49\textwidth,height=0.340\textwidth]{S_mu_proton_vs_costheta.png}
%\caption{ Example distributions of the total muon MC signal at 1000 m ($S_{\mu,i,j}$) as a function of zenith angle. Distributions from TD simulations at $10^{19}$ eV for iron-induced showers obtained with EPOS-LHC (left) and proton-induced showers with QGSJetII-0.4 (right).}
%\label{fig44}
%\end{figure}
The  average fraction of the ground signal induced by muons $g_{\mu,i}(\theta)$ has been  calculated in many analyses. This fraction  depends on the zenith angle  and  primary type, but only slightly on different hadronic interactions models~\cite{balaz}.  Here the $g_{\mu,i}(\theta)$ fractions is obtained from the analysis of muon traces  from  dense stations located at a distance of 1000 m from the shower core for showers with energies of $10^{18.5}-10^{19}$ eV~\footnote{The following parameterizations of the muon fraction were used:
$g_{\mu,\mathrm{p}}(\theta)=0.592896\sec^{5}(\theta)-4.13028\sec^{4}(\theta)+10.8848\sec^{3}(\theta)- 13.3109\sec^{2}(\theta) +7.75816\sec(\theta) -
  1.44004$;
$ g_{\mu,\mathrm{He}}(\theta)=-1.49048\sec^{5}(\theta)+11.1882\sec^{4}(\theta)-33.6204\sec^{3}(\theta) +50.5005 \sec^{2}(\theta)-37.3599\sec(\theta)+11.1662$;
 $ g_{\mu,\mathrm{N}}(\theta)= 0.109708\sec^{5}(\theta)-0.226697 \sec^{4}(\theta)-1.54144\sec^{3}(\theta) +6.01872 \sec^{2}(\theta)-6.85386\sec(\theta) +2.897$; 
$ g_{\mu,\mathrm{Fe}}(\theta)=-1.37866\sec^{5}(\theta)+10.9333 \sec^{4}(\theta)-34.7169\sec^{3}(\theta)+54.8469\sec^{2}(\theta)-42.4105\sec(\theta)
+13.1689$.}.

%In Figure~\ref{fig44} we shown  the  muon signal as a function of the zenith angle calculated for EPOS-LHC (left) and QGSJetII-04 (right). Note that left plot of Figure~\ref{fig4} shows the muon signal (our MC True) present in the MOCK-DATA set, whose average value over the investigated zenith angle range is about 23.1 VEM. We would like to recover this value using TD simulations obtained with the QGSJetII-0.4 model and the method described in this section. 
% 
\vspace{-0.3cm}
\section{Results}
\begin{figure}[t]
\vspace{-0.5cm}
\centering
\includegraphics[width=0.99\textwidth,height=0.34\textwidth]{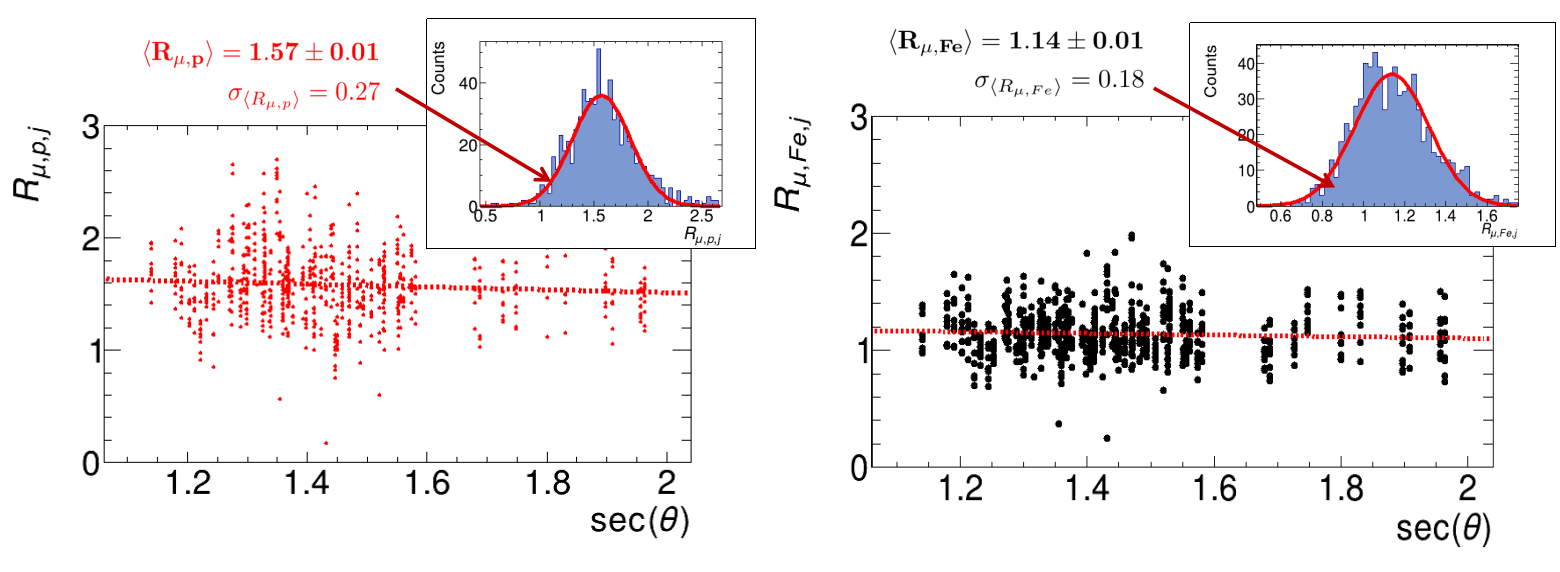}
\vspace{-0.4cm}
\caption{The $R_{\mu,j}$-factor as a function of zenith angle, for $\log_{10}(E/eV)=19$ proton,  and iron induced air showers obtained with QGSJetII-04. The dashed red line indicates a linear fit to the data. Picture inset shows $R_{\mu,i}$ distribution (blue histogram) with the Gaussian fit to the data points (solid red line). From the Gaussian fit, we can obtain the mean value of the  muon correction factor $R_{\mu,i}$ for different primaries, see  Table 2.}
\label{fig4}
\vspace{-0.2cm}
\end{figure}

\vspace{-0.2cm}
The  total muon  signals at 1000 m reconstructed from TD simulations are listed in Table 1  for EPOS-LHC and QGSJETII-04. As we can see, the QGSJetII-04 model always tends to predict smaller SD muon signals than the EPOS-LHC model, which is the expected behavior according to the results presented in~\cite{tang1}. The average number of muons is larger in EPOS-LHC than QGSJetII-04, but  the energy spectrum of the muons at ground is different between the models. There are more muons with lower energies in EPOS-LHC which also leads to a difference in the ground-based muon signal between EPOS-LHC and QGSJetII-0.4. From Table 1,  we can see that the average ratio $r=S_{1000,\mu,i}^{\mathrm{MC-EPOS}}/S^{\mathrm{MC-QGSJet.}}_{1000,\mu,i}$ is about  $1.11\pm 0.04$.  
%% --------------------------------------------------------------------------------
\begin{table}[h]\label{table1}
\small
\caption{\label{ab:ratioTD}Summary of the SD station signals from TD simulations: 
the mean muon signal at 1000 m, $S_{1000,\mu,i}$  and its standard  deviation (st. dev.) for  EPOS-LHC and QGSJetII-04 are listed.}
\begin{tabular}{ccccccc} \hline \hline
  model                &    primary type & $ \langle S_{1000,\mu,i}^{MC} \rangle$   &  st. dev  & model &  $\langle S^{MC}_{1000,\mu,i}\rangle$  & st. dev.\\ 
                  &    $i$            & [VEM]   &  [VEM]  &   &  [VEM]  & [VEM]\\
\hline

  \bf EPOS-LHC  &  \bf p                 & $16.89\pm 0.31$  & 5.1  &   \bf  QGGSJet II -04 & $15.05\pm 0.3$  &4.3 \\
                            & \bf  He    &  $18.74\pm 0.42$ & 5.7   & & $16.82\pm 0.4$ 
                            &  4.7 \\
                            & \bf  N     & $20.67\pm 0.37$  & 5.9       & &$18.96\pm 0.4$  &5.1\\
                            & \bf  Fe    &  $23.09\pm 0.42$ & 6.4     &  & $21.08\pm 0.4$ & 5.7\\  \hline \hline
\end{tabular}
\vspace{-0.2cm}
\end{table}
Figure~\ref{fig4} shows the  muon rescaling factor obtained from Eq.~(4) for different primaries. It can be seen, the average value of $R_{\mu,i}$ depends only slightly on the zenith angle and, as expected, decreases for heavier primaries. This is also confirmed by the values calculated from the Gaussian fit to the histogram of the $R_\mu$ distribution.
In  Table 2, {\it the  total muon signal }$S_{\mu}^{\mathrm{MOCK-DATA}}$ defined~\footnote{This is by  definition from Eqs. (1) and (2).} by: $\langle R_{\mu,i}\rangle \times \langle S^{\mathrm{MC}}_{1000,\mu,i} \rangle$  is listed. The reconstructed muon signal is consistent within 2-4\% with the average muon signal present in our MOCK-DATA set ($S^{\mathrm{MC-True}}_{\mu}=23.09$ VEM)~\footnote{ $S^{\mathrm{MC-True}}_{\mu}$ is known from initial simulations from which MOCK-DATA are obtained, while $S_{\mu}^{\mathrm{MOCK-DATA}}$ is its estimate obtained from TD procedure.}, which shows the correctness of the proposed method.
\begin{table}[h]
\small
\centering
\caption{\label{ab:ratioTD}
The mean value of the muon rescaling parameters $R_{\mu,i}$ calculated from Eq.~(4)  and its standard deviation $\sigma_{ \langle R_{\mu,i} \rangle}$ for different $i$ primaries. Also, the corresponding mean values of the total muon SD signal $S^{\mathrm{MC}}_{1000,\mu,i}$ from QGSJetII-0.4 model, reconstructed muon SD signal at 1000 m expected in the MOCK-DATA set and the ratio $k \equiv (\langle R_{\mu,i}\rangle \times\langle S^{\mathrm{MC}}_{1000,\mu}\rangle-S^{\mathrm{MC-True}}_{\mu})/S^{\mathrm{MC-True}}_{\mu}$ are listed. }
\begin{tabular}{cccccc} \hline \hline
 primary type & $\langle R_{\mu,i} \rangle $  & $\sigma_{ \langle R_{\mu,i} \rangle  }$&   $\langle S^{\mathrm{MC}}_{1000,\mu,i} \rangle$   &  $\langle R_{\mu,i} \rangle \times \langle S^{\mathrm{MC}}_{1000,\mu,i}\rangle$  &  k \\ 
 $i$  &  &  &   [VEM] &   [VEM]  &  [\%] \\
 \hline \hline
 p & $1.57\pm 0.01$   & 0.27 & 15.05   &  23.63  & +2.3 \\
 He & $1.42\pm 0.01$  & 0.26 & 16.82  & 23.88    &       +3.4    \\  
 N & $1.26\pm 0.01$   & 0.21 & 18.96   &23.89     &    +3.5       \\  
 Fe & $1.14\pm 0.01$  & 0.18 &21.08   & 24.00   & +3.9  \\                        
\hline \hline
\end{tabular}
\vspace{-0.4cm}
\end{table}

The number of muons in an air shower is another powerful tracer of the mass of a primary particle. Simulations  and measurements have confirmed that the  number of muons produced, $N_\mu$, rises almost linearly with the primary energy $E$, and increases with a small power of the cosmic-ray mass $A$. This behavior can be understood in terms of the Heitler-Matthews model of hadronic air showers~\cite{heitler}, which predicts $N_{\mu}=A(\frac{E/A}{\epsilon^{\pi}_{c}})^{\beta}=N_{\mu,\mathrm{p}}(E/A)^{1-\beta}$, with $\beta \simeq 0.9$~\footnote{The  $N_{\mu,\mathrm{p}}$ is the number of muons for proton shower
and $\epsilon^{\pi}_{c}$ is the critical energy  at which pions decays into muons.}. Detailed simulations of $\beta$ show further dependencies on hadronic-interaction properties, like the multiplicity,  charge ratio and  baryon anti-baryon pair production~\cite{ulrich}. Thus, measurement of the $\beta$-exponent can effectively constrain the parameters governing hadronic  interactions and improve the accuracy of hadronic models. Assuming that the  average muon signal $S_{\mu}$ is proportional to $ N_{\mu}$ and   calculating the average logarithm of the muon number $N_{\mu,i}$  for $i$ primary and iron (A=56), we  get: $\beta_{i}=1-\frac{\ln \langle S^{\mathrm {MC}}_{\mu,\mathrm {Fe}}\rangle -\ln \langle S^{\mathrm {MC}}_{\mu,i} \rangle }{\ln 56-\ln A_{i}}$. The  calculated $\beta$-exponent for a given interaction model but a different primary, i.e. for the muon signal listed in the Table 1, is about 0.92, which is quite close to the value  reported in~\cite{calzon}, e.g. $\beta=0.927$ for EPOS-LHC and $\beta=0.925$ for QGSJetII-0.4. This cross-check of $\beta$ calculation  is also a validation of our TD simulations. However, we can calculate the $\beta$-exponent  using  the reconstructed muon signal for each primary $i$, e.g. $S^{\mathrm {MOCK-DATA}}_{\mu,i} \equiv \langle R_{\mu,i} \rangle \times \langle S^{\mathrm {MC}}_{1000,\mu,i}\rangle$. In this case, the  exponent $\beta_i$  can be given  by:
$\beta_{i}=1-\frac{\ln \langle S^{\mathrm {DATA-MOCK}}_{\mu,\mathrm Fe}\rangle -\ln \langle S^{\mathrm {DATA-MOCK}}_{\mu,i} \rangle }{\ln 56-\ln A_{i}}=1-\frac{\ln( \langle R_{\mu,\mathrm {Fe}}\rangle \langle S^{\mathrm {MC}}_{\mu,\mathrm {Fe}}\rangle) -\ln (\langle R_{\mu,i}\rangle \langle S^{\mathrm {MC}}_{\mu,i}\rangle) }{\ln 56-\ln A_{i}}$.

\begin{figure}[t]
\vspace{-0.7cm}
\centering
\includegraphics[width=0.49\textwidth,height=0.28\textwidth]{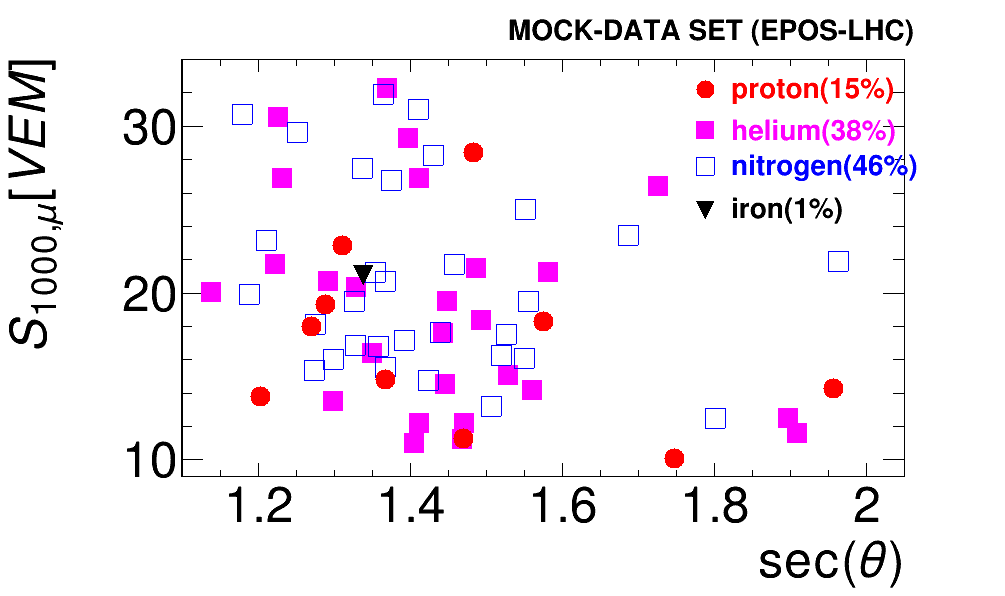}
\includegraphics[width=0.50\textwidth,height=0.28\textwidth]{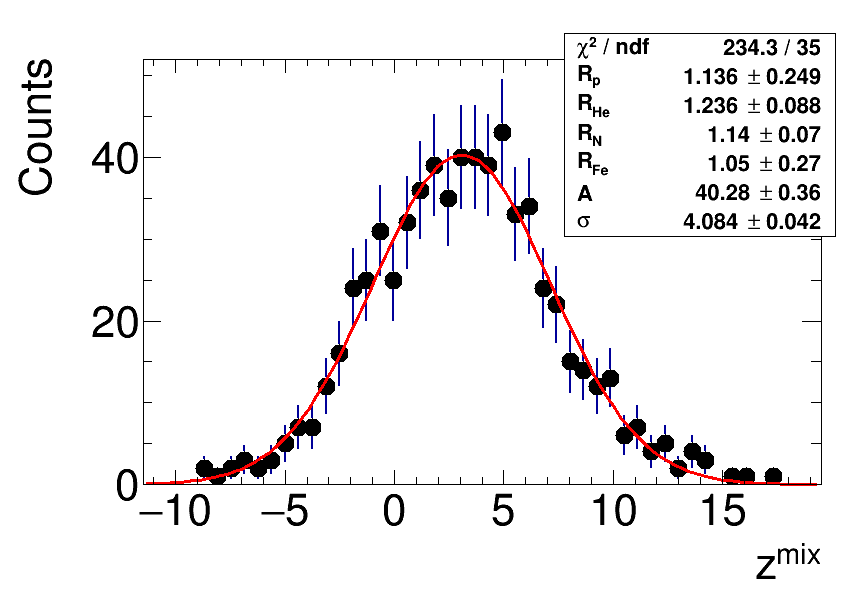}

\includegraphics[width=0.49\textwidth,height=0.33\textwidth]{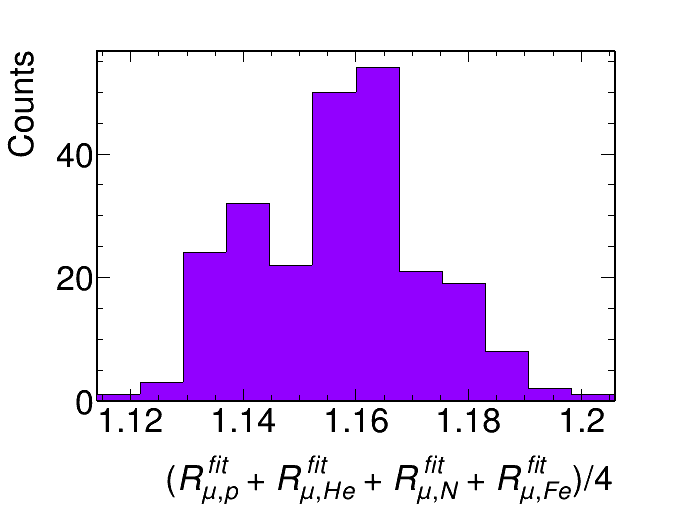}
\includegraphics[width=0.49\textwidth,height=0.33\textwidth]{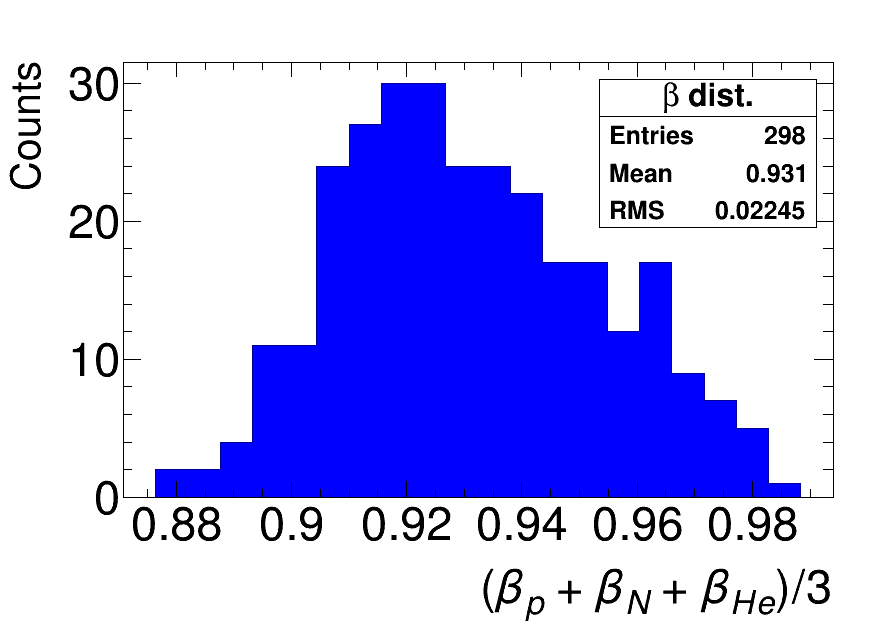}
\caption{ Upper left: The $S_{1000, \mu,i}$   as a function of zenith angle. The number of events (68 events in total) follow the fraction of elements reported by Auger for EPOS-LHC~\cite{belido}; Upper right:  The $z^{\mathrm{mix}}$  distributions  (histogram) for MOCK-DATA set. The red line shows an example of a Gaussian fit given by Eq.~(5); Lower panels: The average muon rescaling  factor (left) and an average $\beta_i$-exponent distribution (right) for events passing our selection criterion are shown.}
\label{fig:z_epos}
\end{figure}

In the following, we show how to compute the $\beta_i$-exponent for a set of hybrid events that consist of a certain fraction of events with different primaries. In the first step, we generate a new MOCK--DATA set, see Figure~\ref{fig:z_epos} (top left). The MOCK--DATA set is constructed in such a way that the considered sample of events corresponds to the fraction of elements ($f_i$) reported by Auger at $10^{19}$ eV and for EPOS-LHC, i.e. $f_{\mathrm{p}}\simeq15\%$, $f_{\mathrm{He}}\simeq38\%$, $f_{\mathrm{N}}\simeq46\%$ and $f_{\mathrm {Fe}}\simeq1\%$~\cite{belido}. Furthermore, the zenith angle distribution is consistent with the corresponding distribution from the TD simulations. It is also worth mentioning that for the MOCK-DATA set, the average muon signal for the primary $i$ is similar, i.e. within $\pm 0.5$ VEM, to the value given in Table 1 for the EPOS-LHC model. In the next step, for a single event we can calculate the $z^{\mathrm{mix}}$-variable defined as: $z^{\mathrm {mix}}_{j} \equiv S^{\mathrm {MOCK-DATA}}_{1000,j}-\sum_{i}f_{i}S^{\mathrm {MC}}_{\mu,i,j} $. The distribution  of the $z^{\mathrm {mix}}$-variable (histogram) is shown
in  Figure~4 (upper right). Finally, to the $z^{\mathrm {mix}}$-histogram we fit the Gaussian function given by:
\begin{eqnarray}
P(A,\sigma,{\bf R^{fit}_{\mu}})=A\exp(-(z^{\mathrm {mix}}-\langle z^{\mathrm {mix}} \rangle )^2/2\sigma^2),\\
  \langle z^{\mathrm{ mix}} \rangle=\sum f_{i}\times \langle S^{\mathrm {MC}}_{\mu,i}\rangle (R^{\mathrm {fit}}_{\mu,i}-1),
\end{eqnarray}
where fitting parameters are amplitude $A$,  the standard deviation $\sigma$ and four rescaling parameters ${\bf R_{\mu}^{\mathrm {fit}}}=\{R^{\mathrm{fit}}_{\mu,\mathrm{p}},R^{\mathrm{fit}}_{\mu,\mathrm{He}},R^{\mathrm{fit}}_{\mu,\mathrm{N}},R^{\mathrm{fit}}_{\mu,\mathrm{Fe}}\}$. Note that, following  Eq.~(3), the mean of the total muon signal will be proportional to the $\langle z^{\mathrm {mix}}\rangle$, and {\it the factor  $R^{\mathrm {fit}}_{\mu,i}\times  \langle S^{\mathrm {MC}}_{\mu,i}\rangle$   is by definition  the  contribution of the  primary $i$ to the total muon signal}. 
In order to find the most likely solutions, we fitted  this Gaussian function to $z^{\mathrm{mix}}$-histogram using the ROOT routine gMinuit~\footnote{In fact, we performed many minimizations with  different initial muon scaling parameters, e.g.   we performed a scan in $R_{\mu,i} = \{1 - 2\}$  with step=0.005.   The correct solutions  should fulfill the criterion:  $ R_{\mu,\mathrm{p}} \langle S_{\mu,\mathrm{p}}^{\mathrm MC}\rangle  <  R_{\mu,\mathrm{He}} \langle S_{\mu,\mathrm{He}}^{MC}\rangle < R_{\mu,\mathrm{N}}\langle S_{\mu,\mathrm{N}}^{\mathrm {MC}}\rangle  <  R_{\mu,\mathrm{Fe}}\langle S_{\mu,\mathrm{Fe}}^{\mathrm {MC}}\rangle$, which is  expected from the physics of extensive air showers  that  muon numbers for lighter elements should be smaller than for heavier  elements. To account for different cosmic-ray compositions derived using EPOS and QGSJet models, an additional cut:  $R^{\mathrm{fit}}_{\mu,\mathrm{He}}>R^{\mathrm{fit}}_{\mu,i}$ where $i \in \{\mathrm{p, N, Fe}\} $ was used.}. 

The results of this procedure are shown in Figure~\ref{fig:z_epos}     
 (lower panels). One can see that the proposed scheme can recover the ratio in the muon signal between EPOS-LHC and QGSJetII-0.4 on average within -5\%, e.g. ratio of MC true is $r=1.11\pm 0.04$ and the average value in Figure~\ref{fig:z_epos} (bottom left) is 1.16 $\pm 0.02$. The observed difference is due to the fact that the muon signal is slightly different for the MOCK-DATA set than the one listed in Table 1 for EPOS-LHC. We can also recover an average parameter $\beta=0.92$ for the studied system, which is a consequence of the good recovery (less than 6\% on average) of the muon signal for each primary. 
\vspace{-0.3cm}    
\section{Conclusion}
\vspace{-0.2cm}
This paper presents a method for determining muon scaling factors. The method has been applied to the TD reconstruction results of hybrid showers with an average energy of $10^{19}$ eV, for two interaction models:  EPOS LHC and QGSJetII-04, and four primary particle types: proton, helium, nitrogen, and iron. The method enables testing hadronic interaction models, to calculate the average muon signal for a given set of hybrid events, and gives a possibility to calculate the $\beta$-exponent describing the dependence of the number of muons as a function of the primary mass. 

{\bf Acknowledgments:} The  authors  are very grateful to the Pierre Auger Collaboration for providing the simulations for this contribution. We would like also to thank Kevin Almeida Cheminant for various cross-checks of the presented analysis~\footnote {\small  We want also to acknowledge support in Poland from National Science Centre grant No. 2016/23/B/ST9/01635, grant No. 2020/39/B/ST9/01398 and from the Ministry of Science and Higher Education grant No. DIR/WK/2018/11.}.
\vspace{-0.4cm}
\let\oldbibliography\thebibliography
\renewcommand{\thebibliography}[1]{\oldbibliography{#1}
\setlength{\itemsep}{1.6pt}}

\end{document}